\documentclass[superscriptaddress,prl]{revtex4}
\usepackage{amsmath}
\usepackage{graphicx}

\begin{document}

\title{Radiative Electroweak Symmetry-Breaking Revisited}
\author{V.\ Elias}
\affiliation{Perimeter Institute for Theoretical Physics, 35 King Street North, Waterloo,
ON, N2J 2W9, Canada}
\affiliation{Department of Applied Mathematics, The University of Western Ontario,
London, ON, N6A 5B7, Canada}
\author{R.B.\ Mann}
\affiliation{Perimeter Institute for Theoretical Physics, 35 King Street North, Waterloo,
ON, N2J 2W9, Canada}
\affiliation{Department of Physics, University of Waterloo, Waterloo, ON, N2L 3G1, Canada }
\author{D.G.C.\ McKeon}
\affiliation{Department of Applied Mathematics, The University of Western Ontario,
London, ON, N6A 5B7, Canada}
\author{T.G. Steele}
\affiliation{Department of Physics \& Engineering Physics, University of Saskatchewan,
Saskatoon, SK, S7N 5E2, Canada }

\begin{abstract}
In the absence of a tree-level scalar-field mass, renormalization-group
methods permit the explicit summation of leading-logarithm contributions to
all orders of the perturbative series within the effective potential for $%
SU(2)\times U(1)$ electroweak symmetry. This improvement of the effective
potential function is seen to reduce residual dependence on the
renormalization mass scale. The all-orders summation of leading logarithm
terms involving the dominant three couplings contributing to radiative
corrections is suggestive of a potential characterized by a plausible Higgs
boson mass of $216$ GeV. However, the tree potential's local minimum at $%
\phi =0$ is restored if QCD is sufficiently strong.
\end{abstract}

\maketitle

Over thirty years ago, S.\ Coleman and E.\ Weinberg \cite{1} demonstrated
how spontaneous symmetry breaking may occur through radiative corrections to
a conformally-invariant Lagrangian in which no quadratic mass term appears.
Such symmetry breaking, in which the scalar-field vacuum expectation value $%
\langle \phi \rangle $ is the only source of scale, is of particular
relevance for the spontaneous breakdown of $SU(2)\times U(1)$ electroweak
symmetry, which necessarily requires a mechanism within any embedding theory
to keep any such quadratic mass term minimally contaminated by the
unification mass scale--- the absence of such a mass term implies that this
mechanism is exact \cite{sher}. We emphasize that such a mechanism, whether
exact or nearly so, is a necessary component of the Standard Model, though
the nature of this mechanism (possibly conformal invariance) remains
unknown. In the absence of an explicit scalar-field mass term (i.e. the
``exact mechanism''), the one-loop (1L) effective potential for $SU(2)\times
U(1)$ gauge theory is given by \cite{1}
\begin{equation}
V_{eff}^{(1L)}=\frac{\lambda \phi ^{4}}{4}+\phi ^{4}\left[ \frac{12\lambda
^{2}-3g_{t}^{2}}{64\pi ^{2}}+\frac{3(3g_{2}^{4}+2g_{2}^{2}g^{\prime
\;2}+g^{\prime \;4})}{1024\pi ^{2}}\right] \left( \log \frac{\phi ^{2}}{\mu
^{2}}-\frac{25}{6}\right) .  \label{eq1}
\end{equation}%
There are four distinct coupling constants appearing in Eq.\ (\ref{eq1}),
the $SU(2)$ coupling constant $g_{2}$, the $U(1)$ coupling constant $%
g^{\prime }$, the $t$-quark Yukawa coupling constant $g_{t}$, and the
quartic scalar-field self-interaction coupling constant $\lambda $. The
radiative symmetry breaking scenario of Ref.\ \cite{1}, which preceded the
discovery of the massive top quark, led to a value for $\lambda $
proportional to $g_{2}^{4}$ and a scalar field mass of order $10$ GeV. The
presence of a large Yukawa couplant [$g_{t}^{2}\simeq 1.0>>g_{2}^{2},\;{%
g^{\prime }}^{2}$] spoils this scenario; the ${\mathcal{O}}(g_{t}^{2})$
value of $\lambda $ required for radiative symmetry breaking would be so
large that subsequent leading logarithm terms [\textit{e.g.} $\lambda
^{3}\phi ^{4}\log ^{2}(\phi ^{2}/\mu ^{2})$] would be too large to neglect.

In the present work, we explicitly sum all such leading logarithm terms
within the full perturbative series for the effective potential \cite{1a} to
examine the viability of radiative electroweak symmetry breaking. We find
the potential is minimized for a Higgs mass of $216$ GeV, and observe some
evidence that this value may be stable after including contributions from
subsequent-to-leading logarithms.

If we denote the dominant Standard Model couplants as $x\equiv
g_{t}^{2}/4\pi ^{2}$, $y\equiv \lambda /4\pi ^{2}$, and $z\equiv
g_{3}^{2}/4\pi ^{2}$ [QCD contributes to leading logarithms past one-loop
order], which are much larger than corresponding couplants for $g_{2}$, $%
g^{\prime }$ and non-$t$-quark Yukawa couplings, this series is of the form $%
V_{eff}=\pi ^{2}\phi ^{4}\sum C_{n,k,\ell ,p}x^{n}y^{k}z^{\ell }L^{p}$,
where $L(\mu )\equiv \log [\phi ^{2}(\mu )/\mu ^{2}]$. The leading
logarithms in this series are those terms one degree lower in the power of
the logarithm $L$ than in the aggregate power of the couplants $\left\{
x,y,z\right\} $, 
\begin{equation}
V_{LL}=\pi ^{2}\phi ^{4}S_{LL}=\pi ^{2}\phi ^{4}\left\{ \sum_{n=0}^{\infty
}x^{n}\sum_{k=0}^{\infty }y^{k}\sum_{\ell =0}^{\infty }z^{\ell }C_{n,k,\ell
}L^{n+k+\ell -1}\right\} ,\;\;(C_{0,0,0}=0).  \label{eq2}
\end{equation}%
The series $S_{LL}$ is determined entirely by one-loop contributions to the
renormalization-group (RG) equation, \textit{i.e.}, by those contributions
that either lower the power of $L$ by one or raise the aggregate power of
the couplants by one \cite{2,sher}: 
\begin{equation}
\left[ -2\frac{\partial }{\partial L}+\left( \frac{9}{4}x^{2}-4xz\right) 
\frac{\partial }{\partial x}+\left( 6y^{2}+3yx-\frac{3}{2}x^{2}\right) \frac{%
\partial }{\partial y}-\frac{7}{2}z^{2}\frac{\partial }{\partial z}-3x\right]
S_{LL}(x,y,z,L)=0.  \label{eq3}
\end{equation}%
In eq. (\ref{eq3}), the coefficients of $\frac{\partial }{\partial x}$, $%
\frac{\partial }{\partial y}$, $\frac{\partial }{\partial z}$ are
respectively the one-loop beta-functions for $x,y,z$ (where $\beta _{x}=\mu 
\frac{dx}{d\mu }$); the final term in eq. (\ref{eq3}) is four times the
one-loop scalar field anomalous dimension. For example, the leading
coefficients $C_{0,1,0}=1$, $C_{1,0,0}=C_{0,0,1}=0$, follow from the $%
\lambda \phi ^{4}/4$ tree-order potential. Upon substitution of Eq.\ (\ref%
{eq2}) into Eq.\ (\ref{eq3}), one easily sees that $C_{0,2,0}=3$, $%
C_{2,0,0}=-3/4$, and that the remaining four degree-2 coefficients $%
C_{i,j,2-i-j}$ equal zero, leading to a recovery of the $\left\{ \lambda
^{2},g_{t}^{2}\right\} $-contributions to the potential (\ref{eq1}).

We find it convenient to express the series (\ref{eq2}) in the form 
\begin{equation}
S_{LL}=yF_{0}(w,\zeta )+\sum_{n=1}^{\infty }x^{n}L^{n-1}F_{n}(w,\zeta )
\label{eq4}
\end{equation}%
where $w\equiv 1-3yL$ and $\zeta \equiv zL$, and where 
\begin{equation}
F_{n}(w,\zeta )=\sum_{\ell =0}^{\infty }\sum_{k=0}^{\infty }C_{n,\ell
,k}\left( \frac{1-w}{3}\right) ^{\ell }\zeta ^{k}\equiv
\sum_{k=0}^{n+1}f_{n,k}(\zeta )\left[ \frac{w-1}{w}\right] ^{k}.  \label{eq5}
\end{equation}%
By using Eq.\ (\ref{eq3}) to obtain sequential partial differential
equations relating $F_{0}=1/w$ to $F_{1}(w,\zeta )$ and $F_{2}(w,\zeta )$,
we are able to solve explicitly for these quantities. For $p\geq 3$ one can
show from Eq.\ (\ref{eq3}) that 
\begin{equation}
\begin{split}
0=& \left( \left[ (7\zeta ^{2}/2)\frac{d}{d\zeta }+4p\zeta \right] +\left[
2\zeta \frac{d}{d\zeta }+2(p-1)+2k\right] \right) f_{p,k}(\zeta ) \\
& -\left[ (9p-21)/4+3k\right] f_{p-1,k}(\zeta )+3(k-1)f_{p-1,k-1}(\zeta ) \\
& -\left[ 9(k-1)/2\right] f_{p-2,k-1}(\zeta )+9kf_{p-2,k}(\zeta )-\left[
9(k+1)/2\right] f_{p-2,k+1}(\zeta ),
\end{split}
\label{eq12}
\end{equation}%
where $f_{p,k}\equiv 0$ when $k<0$ or $k>p+1$, and where $f_{p,k}(0)$ is
finite.

We now examine possible radiative spontaneous symmetry breaking for the
RG-improved effective potential $V_{eff}(\phi )=\pi ^{2}\phi ^{4}(S_{LL}+K)$%
, where $K$ is a finite $\phi ^{4}$ counterterm coefficient. As in Ref.\ %
\cite{1}, we choose $\mu =\langle \phi \rangle =v$ [$L=\log \left( \phi
^{2}(v)/v^{2}\right) $], in which case $V_{eff}^{\prime }(v)=0$. The
counterterm $K$ facilitates the fourth-derivative renormalization condition $%
V_{eff}^{(4)}(v)=V_{tree}^{(4)}(v)=24\pi ^{2}y\;(=6\lambda )$. For any
one-loop effective potential of the form $V_{eff}^{(1L)}(\phi )=\pi ^{2}\phi
^{4}\left[ y+\sigma L+K\right] $, this fourth-derivative condition ensures
that $K=-25\sigma /6$, as in Eq.\ (\ref{eq1}). For our RG-improved
potential, we find it convenient to expand $S_{LL}$ in powers of the
logarithm $L$: 
\begin{equation}
S_{LL}=y+B\log {\left( \frac{\phi ^{2}}{v^{2}}\right) }+C\log ^{2}{\left( 
\frac{\phi ^{2}}{v^{2}}\right) }+D\log ^{3}{\left( \frac{\phi ^{2}}{v^{2}}%
\right) }+E\log ^{4}{\left( \frac{\phi ^{2}}{v^{2}}\right) }+...,
\label{eq14}
\end{equation}
We then obtain from Eqs.\ (\ref{eq3}) and (\ref{eq12}) an exact
determination of terms up to ${\mathcal{O}}(L^{4})$ within Eq.\ (\ref{eq14}%
): 
\begin{gather}
B=3y^{2}-\frac{3}{4}x^{2},  \label{eq21} \\
C=9y^{3}+\frac{9}{4}xy^{2}-\frac{9}{4}x^{2}y+\frac{3}{2}x^{2}z-\frac{9}{32}%
x^{3},  \label{eq22} \\
D=27y^{4}+\frac{27}{2}xy^{3}-\frac{3}{2}xy^{2}z+3x^{2}yz-\frac{225}{32}%
x^{2}y^{2}-\frac{23}{8}x^{2}z^{2}+\frac{15}{16}x^{3}z-\frac{45}{16}x^{3}y+%
\frac{99}{256}x^{4},  \label{eq23} \\
\begin{split}
E=& 81y^{5}+\frac{243}{4}xy^{4}-9xy^{3}z+\frac{45}{32}xy^{2}z^{2}-\frac{69}{%
16}x^{2}yz^{2}-\frac{135}{8}x^{2}y^{3}+\frac{531}{64}x^{2}y^{2}z \\
& +\frac{345}{64}x^{2}z^{3}-\frac{603}{256}x^{3}z^{2}+\frac{207}{32}x^{3}yz-%
\frac{8343}{512}x^{3}y^{2}-\frac{459}{512}x^{4}z+\frac{135}{512}x^{4}y+\frac{%
837}{1024}x^{5}.
\end{split}
\label{eq24}
\end{gather}

The procedure for obtaining the Higgs mass, as described below, is
insensitive to any terms in the leading-logarithm series (\ref{eq14}) past $%
\mathcal{O}(L^{4})$. Consequently, Eqs.\ (\ref{eq21})--(\ref{eq24}) are
sufficient to determine the entire leading logarithm contribution to the
scalar field mass to \emph{all} (contributing) orders in $\{x,y,z\}$. These
equations are also obtainable via the method-of-characteristics methodology
of Bando \textit{et al} \cite{bando}, which has been implemented in
conventional (non-radiative) Standard Model symmetry breaking by Quiros and
collaborators \cite{quiros}. The conditions $V_{eff}^{\prime }(v)=0$ and $%
V_{eff}^{(4)}(v)=24\pi ^{2}y$ respectively imply that $y=-B/2-K$ and $K=-[%
\frac{25}{6}B+\frac{35}{3}C+20D+16E]$. Given the phenomenological Standard
Model values for the vacuum expectation value $v=246\,GeV$, the $t$-quark
Yukawa couplant $x(v)=1/4\pi ^{2}$, the QCD couplant $z(v)=\alpha
_{s}(v)/\pi =0.0329$, as evolved from $\alpha _{s}(M_{z})=0.120$ \cite{3},
we find these constraints taken together constitute a degree-5 equation for
the scalar-field self-interaction couplant $y$. The only real positive-$y$
solution that yields a positive second derivative (hence, a local minimum)
is $y=0.0538$. Once $y$ is determined, then $B$, $C$, $D$ and $E$ are also
numerically determined.
To present order, we can approximate the Higgs field propagator pole with the second derivative of
the effective potential at $\phi=v$.
One then finds $m_\phi^2\cong V_{eff}^{\prime \prime }(v)={8}\pi
^{2}v^{2}(B+C)=(216$ $\mathrm{{GeV})^{2}}$.

In assessing the viability of this result, it is of interest to consider
what one would similarly obtain from the one-loop effective potential
augmented by a $\pi ^{2}\phi ^{4}K$ counterterm. Such a potential is seen to
correspond to Eq.\ (\ref{eq14}) with $B$ as given by Eq.\ (\ref{eq21}), but
with $C=D=E=0$. The conditions $V^{\prime }(v)=0$, $V^{(4)}(v)=24\pi ^{2}y$
are then seen to lead to a solution $y=0.093$, $m_{\phi }=350$ GeV. Such a
mass is well outside the $\mathcal{O}(200\,\mathrm{GeV})$ bound on $m_{\phi }
$ from corrections to electroweak theory \cite{3}. Moreover, it is easy to
demonstrate that this value for $y$ is too large to be meaningful. The
contributions of $y$ alone to the $\beta $-function for its own evolution
correspond to the $\beta $-function of an ${O}(4)$ symmetric scalar field
theory, which is known to five-loop order \cite{4}: 
\begin{equation}
\lim_{\overset{x\rightarrow 0}{z\rightarrow 0}}\mu \frac{dy}{d\mu }=6y^{2}-%
\frac{39}{2}y^{3}+187.85y^{4}-2698.3y^{5}+47975y^{6}+...  \label{eq28}
\end{equation}%
If $y=0.093$, terms of this series increase after the second term,
indicative of a failure to converge. By contrast, terms of the series (\ref%
{eq28}) decrease if $y=0.0538$. Similar results characterize this same scalar
field theory's anomalous dimension, whose terms decrease monotonically when $%
y=0.0538$, but fail to do so when $y=0.093$. Of course it is of greater
interest to estimate possible corrections to our result to 2-loop order. For
our parameter values $x=0.0253$, $y=0.0538$, $z=0.0329$, the two loop
contributions to the Standard Model beta functions and anomalous scalar-field dimension
provide corrections no larger than $17\%$\ of their one-loop counterparts.
This provides us with further confidence that the $216\,$GeV Higgs mass
prediction will be stable upon summation of subsequent next-to-leading
logarithms. 

\begin{figure}[tbh]
\includegraphics[scale=0.4]{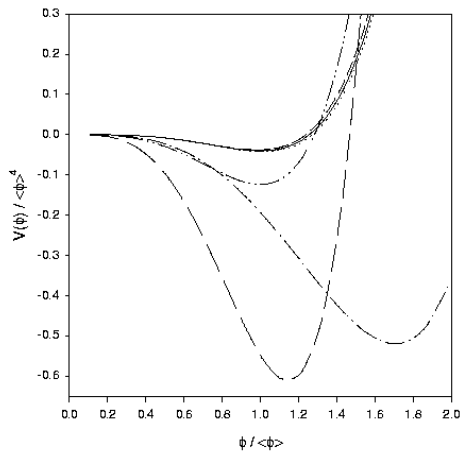}
\caption{Residual scale dependence of the standard model effective potential
with (upper three curves) and without (lower three curves) summation of
leading logarithms, as discussed in the text. For the resummed curves, the
solid line represents $\protect\mu =v$, the dashed line represents $\protect%
\mu =v/2$, and the dotted line represents $\protect\mu =2v$. For the
unsummed curves, the dashed-double-dotted curve represents $\protect\mu =v$,
the dashed-single-dotted curve represents $\protect\mu =v/2$, and the dashed
curve is for $\protect\mu =2v$. }
\label{scale_fig}
\end{figure}

In Figure \ref{scale_fig} we compare the residual scale- ($\mu $-)
dependence of $V_{eff}(\phi )=\pi ^{2}\phi ^{4}\left( S_{LL}+K\right) $
obtained via Eqs.\ (\ref{eq21})--(\ref{eq24}) to that of the one-loop
effective potential discussed in the preceding paragraph. Such dependence in
both potentials occurs explicitly through $L(\mu )$ and implicitly through
couplants whose one-loop evolution in $\mu $ is anchored to the $\mu =v$
initial values given above (\textit{e.g.} $x(v)=1/4\pi ^{2}$). The $K\phi
^{4}$ counterterms in both potentials are each assumed to be RG-invariant ($%
K(v)\phi ^{4}(v)$), since the subleading logarithm contributions ultimately
devolving from such terms are uncontrolled by Eq.\ (\ref{eq3}). For $\mu
=\{v/2,v,2v\}$, the curves exhibit the dependence of the potentials on the
RG-invariant initial value $\phi (v)$ for the evolution of $\phi (\mu )$ [$%
(\mu /\phi )d\phi /d\mu =-3x(\mu )/4$]. Figure \ref{scale_fig} shows that
summation of leading logarithms substantially reduces the residual scale
dependence of the effective potential. Moreover, if we assume such scale
dependence to be indicative of next-order corrections, we can expect only
modest departures from the $m_{\phi }=216\,\mathrm{GeV}$ prediction at $\mu
=v$: $m_{\phi }$ varies from 208 GeV at $\mu =v/2$ to 217 GeV at $\mu =2v$.
We find such uncertainties in $m_{\phi }$ to dominate over much smaller ones
deriving from (Standard-Model) uncertainties in the couplant values $x(v)$
and $z(v)$. We have also verified by numerical calculation of the RG
equations that \ the method-of-characteristics methodology of Bando et al %
\cite{bando,quiros} yields the same results for the effective potential and
the Higgs mass to within a value of 0.2\%.

Note also that the scalar field mass of order $216\, \mathrm{GeV}$ we obtain
from the aggregate contribution of leading logarithms to the
purely-radiative breakdown of $SU(2) \times U(1)$ electroweak symmetry is
accompanied by a scalar-field interaction couplant $y = 0.0538$
substantially larger than that anticipated from conventional spontaneous
symmetry breaking (deriving from a potential with an initially-negative
quadratic term), in which a $216\, \mathrm{GeV}$ Higgs particle would
necessarily correspond to a value $y=\lambda/4\pi^2 =
m_\phi^2/(8\pi^2\langle\phi\rangle^2 = 0.0097$. If electroweak
symmetry-breaking is purely radiative, then y-sensitive processes such as
the $W^+W^-\to ZZ$ scattering cross-section \cite{nierste} will necessarily
be larger than anticipated from conventional spontaneous symmetry breaking.
Consequently, if an $\mathcal{O}(200 \,\mathrm{GeV})$ Higgs were discovered,
a clear signal of radiative symmetry breaking would be a corresponding
order-of-magnitude-or-more enhancement of $\sigma\left(W^+W^-\to ZZ\right)$
over the value expected from such a Higgs mass.

One of the motivations for summing leading logarithms is to ascertain the
negative large-logarithm behaviour of the effective potential, behaviour
corresponding to the zero-field limit of the potential. We do not consider
positive large-logarithm behaviour because of the intervening Landau
singularity at $w=0$ [Eqs. (\ref{eq4}) and (\ref{eq5})], corresponding to $%
\phi (v)\approx 22v$ \cite{6ex}. When $|L|$ is very large, we find that $%
yF_{0}\rightarrow -1/3L$, $xF_{1}\rightarrow 2(x/z)/L$, and $%
x^{2}LF_{2}\rightarrow -3(x^{2}/z^{2})/2L$. The large-$|L|$ behaviour of
subsequent terms in the series (\ref{eq4}) can be extracted by noting that
the first term on the RHS of Eq.\ (\ref{eq12}) dominates the second term
when the magnitude of $\zeta (=zL)$ is large, and that $F_{p}(w,\zeta )\sim
\sum_{k=0}^{p+1}f_{p,k}(\zeta )$ in this large-$|L|$ limit $\left[ \left( 
\frac{w-1}{w}\right) \sim 1\right] $. One then finds after a little algebra
that when $|\zeta |$ is large, 
\begin{equation}
\left[ (7\zeta ^{2}/2)\frac{d}{d\zeta }+4p\zeta \right] F_{p}=\frac{9p-21}{4}%
F_{p-1},\;\;\;p\geq 3.  \label{eq29}
\end{equation}%
In the large $|\zeta |$ limit, we find that $F_{2}\sim
\sum_{k=0}^{3}f_{2,k}(\zeta )\sim (-3/2)\zeta ^{-2}$. Eq.\ (\ref{eq29})
implies that $F_{p}\sim f_{p}\zeta ^{-p}$, where the numerical factors $f_{p}
$ follow from $f_{2}=-3/2$ via the recursion relation $%
f_{p}=(9p-21)f_{p-1}/2p$. Note that $x^{p}L^{p-1}F_{p}(w,\zeta )\sim \left( 
\frac{x}{z}\right) ^{p}f_{p}/L$ in the large-$|L|$ limit; each term in the
series (\ref{eq4}) is inversely proportional to $L$ when $|L|$ is large.
Moreover, if $(x/z)<2/9$, the above recursion relation for $f_{p}$ can be
utilized to obtain the closed-form series summation $S_{LL}\sim -\left( 
\frac{1}{3L}\right) \left( 1-\frac{9x}{2z}\right) ^{4/3}$. For sufficiently
strong QCD, this result implies that a Standard Model effective potential
based upon a massless tree potential exhibits a local \emph{minimum}, rather
than a maximum, at $\phi =0$ (\textit{i.e.} $L\rightarrow -\infty $). Such a
conclusion, however, does not follow if $x/z$ is outside its radius of
convergence (\textit{i.e.} if $x/z>2/9$), as is the case for the empirical
standard model [$x(v)\simeq 1/4\pi ^{2}$, $z(v)=\alpha _{s}(v)/\pi \simeq
0.033$].

Recent work \cite{5} based upon Pad\'{e} approximants constructed from the
QCD $\beta $-function series suggests for up to five light flavours that the
QCD couplant may exhibit the same double-valued behaviour known to
characterize $N=1$ supersymmetric Yang-Mills (SYM) theory, in which
coexisting strong-couplant and (asymptotically-free) weak-couplant phases
evolve toward a common infrared attractive point \cite{6}. If the strong
phase is sufficiently strong ($x/z<2/9$), one can envision a scenario in
which the $\phi =0$ local minimum of preserved $SU(2)\times U(1)$ symmetry
is upheld by the strong phase of QCD, but is transformed into a symmetry
breaking minimum at \textsf{$\phi =v$ } if QCD is in its weak phase. Since
for the latter case the minimum at $\phi =v$ $(V(v)<0)$ is deeper than the $%
\phi =0$ $(V(0)=0)$ minimum occurring when QCD is in its strong phase, the
weak phase of QCD is seen to be the preferred one. Thus, if QCD is
characterized by two coexisting phases, as is the case for SYM \cite{6}, the
asymptotic freedom of QCD may be linked to the radiative breakdown of
electroweak symmetry.

\begin{acknowledgments}
We are grateful to V.A.\ Miransky for useful discussions, and to the Natural
Sciences \& Engineering Research Council of Canada (NSERC) for financial
support.
\end{acknowledgments}


\begin{thebibliography}{99}
\bibitem{1} S. Coleman and E. Weinberg, Phys. Rev. D 7, 1888 (1973).

\bibitem{sher} M.\ Sher, Phys.\ Rep.\ 179, 273 (1989).

\bibitem{1a} Such an analysis is anticipated in a $\phi ^{4}$ model in B.\
Kastening, Phys.\ Lett.\ B 283, 287 (1992).

\bibitem{2} T. P. Cheng, E. Eichten and L.-F. Li, Phys. Rev. D9, 2259
(1974); M. B. Einhorn and D. R. T. Jones, Nucl. Phys. B 211, 29 (1983); M.
J. Duncan, R. Philippe and M. Sher, Phys. Lett. B 153, 165 (1985).

\bibitem{bando} M.\ Bando, T.\ Kugo, N.\ Maekawa and H.\ Nakano, Phys.\
Lett.\ B 301, 83 (1993).

\bibitem{quiros} J.A.\ Casas, J.R.\ Espinosa, and M.\ Quiros, Phys.\ Lett.\
B 342, 171 (1995); M.\ Quiros in \textit{Perspectives on Higgs Physics II},
G.L.\ Kane, ed.\ (World Scientific, Singapore, 1997) 148.

\bibitem{3} Particle Data Group [K. Hagiwara \textit{et. al.}], Phys. Rev. D
66, 1 (2002).

\bibitem{4} H. Kleinert, J. Neu, V. Schulte-Frohlinde, K. G. Chetyrkin, S.
A. Larin, Phys. Lett. B 272, 39 (1991); B 319, 545 (E) (1993).

\bibitem{nierste} U.\ Nierste and K.\ Riesselmann, Phys.\ Rev.\ D 53, 6638
(1996).

\bibitem{6ex} A similar ultraviolet singularity can be shown to characterize
the summation of leading logarithms in $V_{eff}$ for massless scalar-field
electrodynamics, a theory for which radiative spontaneous symmetry breaking
is well-established.

\bibitem{5} F. A. Chishtie, V. Elias, V. A. Miransky and T. G. Steele, Prog.
Theor. Phys. 104, 603 (2000).

\bibitem{6} I. I. Kogan and M. Shifman, Phys. Rev. Lett. 75, 2085 (1995);
see also V. Elias, J. Phys. G27, 217 (2001) regarding D.R.T. Jones, Phys.
Lett. B123, 45 (1983).
\end{thebibliography}
\end{document}